\def\bold#1{\setbox0=\hbox{$#1$}%
     \kern-.025em\copy0\kern-\wd0
     \kern.05em\copy0\kern-\wd0
     \kern-.025em\raise.0433em\box0 }
\def\slash#1{\setbox0=\hbox{$#1$}#1\hskip-\wd0\dimen0=5pt\advance
       \dimen0 by-\ht0\advance\dimen0 by\dp0\lower0.5\dimen0\hbox
         to\wd0{\hss\sl/\/\hss}}
\documentstyle[12pt]{article}

\newlength{\dinwidth}
\newlength{\dinmargin}
\newcommand{\resection}[1]{\setcounter{equation}{0}\section{#1}}

\begin{document}
\def\lq{\left [}
\def\rq{\right ]}
\def\LL{{\cal L}}
\def\VV{{\cal V}}
\def\AA{{\cal A}}

\newcommand{\be}{\begin{equation}}
\newcommand{\ee}{\end{equation}}
\newcommand{\bea}{\begin{eqnarray}}
\newcommand{\eea}{\end{eqnarray}}
\newcommand{\nn}{\nonumber}
\newcommand{\dd}{\displaystyle}

\thispagestyle{empty}
\vspace*{4cm}
\begin{center}
  \begin{Large}
  \begin{bf} HIGGS SEARCH BY NEURAL NETWORKS AT LHC\\
  \end{bf}
  \end{Large}
  \vspace{1cm}
  \begin{large}
P. Chiappetta\\
  \end{large}
{\it Centre de Physique Th\'eorique, C.N.R.S. Luminy, France}\\
  \vspace{8mm}
  \begin{large}
P.Colangelo$^1$, P. De Felice$^{1,2}$, G. Nardulli$^{1,2}$\\
  \end{large}
$^1${\it I.N.F.N., Sezione di Bari}\\
$^2${\it Dipartimento di Fisica, Universit\'a
di Bari, Italy}\\
  \vspace{8mm}
  \begin{large}
G. Pasquariello\\
  \end{large}
{\it Istituto Elaborazione Segnale Immagini, C.N.R., Bari, Italy}
  \vspace{5mm}
\end{center}
  \vspace{2cm}
\begin{center}
CPT-93/PE 2969 \\
BARI-TH/159-93 \\
November 1993\\
\end{center}
\vspace{1cm}
\begin{quotation}
\begin{center}
  \begin{Large}
  \begin{bf}
  ABSTRACT
  \end{bf}
  \end{Large}
\end{center}
  \vspace{5mm}
\noindent
We show that neural network classifiers can be used to discriminate Higgs
production from background at LHC for $ 150< M_H<200$ GeV. The results
compare favourably with ordinary multivariate analysis.
\end{quotation}
\newpage
\setcounter{page}{1}
\resection{Introduction}
\par
Experimental data accumulated so far, and especially the LEP results,
strongly support the Standard Model (hereafter denoted as SM) as the
theory of the fundamental interactions at the presently
available energies. Nevertheless the verification
of its validity has to be completed since the top quark and Higgs
boson have not been discovered yet.

As well known, the value of the
Higgs mass is not predictable, but there are indications, arising from the
limits of applicability of the perturbation theory or violation of unitarity,
that it should not exceed
800 GeV \cite{lus}. If Higgs particles below $\approx 1$ TeV are not
discovered, other strong forces could be at work, as predicted by the
Technicolor \cite{tec} scheme, which however, at least in its minimal version,
is not
favored by LEP data \cite{alt}; in this and
other similar approaches the strongly interacting
scalar sector might be revealed by the presence of new vector
bosons \cite{cas} and some light on the electroweak symmetry breaking
could be shed by longitudinal boson scattering. Another theoretical extension
of SM is provided by Supersymmetry (SUSY) (for reviews see\cite{sus}),
which, as
well known, naturally solves the {\it hierarchy problem}
because boson and fermion loops contributions to scalar masses have
opposite signs in SUSY and tend to cancel out, thus avoiding
Higgs masses of the order of the Grand Unification scale $M_{GUT}\sim 10^{16}$
GeV. In the sequel, however, we shall consider only the SM Higgs also
because its search is the first motivation for the future
high energy hadron colliders, and also since one of the SUSY Higgs particles
exhibits a similar behaviour.
\par
The present upper limit on the SM Higgs mass coming from LEP is $63$ GeV
\cite{lep}. Therefore, since
the LEP 200 discovery limit is around $80-90$ GeV, more energetic colliders
are mandatory to pin down the mechanism of the electroweak symmetry
breaking.
\par
In this letter we will consider experiments aiming to discover the SM Higgs at
the future Large Hadron Collider (LHC) which is planned for the end of this
century at CERN.
Within the Standard Model
the discovery of the Higgs particle should
be complicated
by the presence of huge backgrounds. Considerable effort has been
provided by the Aachen workshop of the LHC study groups to clarify this issue,
 and we refer
the interested reader to the Proceedings \cite{lhc} of that Conference for
a comprehensive survey. Our aim here is to analyze
the possibility to use a
neural network (NN) classifier as a tool for a better discrimination
between signal and  background and to evaluate the relative performance
between the neural trigger and traditional statistical methods such as
the multivariate analysis. Given the limits of the present
work we shall not consider the whole Higgs mass range nor we
study all the possible Higgs decay channels, but we shall limit ourselves
to some specific case studies. More precisely we shall analyze
the Higgs mass range $150 - 200$  GeV
and study the  decay into four muons,
which, as shown by the above mentioned LHC  study groups, seems to be the
most favourable decay channel for Higgs discovery.

We first discuss in section 2 a possible choice of the
physical observables useful for
the separation of the Higgs signal from background;
these observables are the input variables
for the NN classifier that is described in the same section.
We present our results and discuss the relative performance
of the NN method and multivariate analysis in section 3. Finally,
in section 4 we draw our conclusions.
\bigskip
\resection{Physical observables and the neural network}
At hadron colliders the dominant mechanism for Higgs production, in the
intermediate mass range we are interested in, is gluon-gluon
fusion. The best decay channel for identification is two real
$Z^0$ bosons for Higgs mass $M_H \ge 2 M_Z$
or, for $M_H \le 2 M_Z$, one real and one off-shell $Z^0$,
followed by their leptonic decays. LHC studies have shown that this
channel is the most efficient one for $130 \le M_H \le 800$ GeV; we shall
consider here the case $M_H\, = \, 150$ GeV where the presence
of one virtual
$Z^0$ renders the analysis more demanding and the case
$M_H\, = \, 200$ GeV just above the threshold for production of
two real $Z$'s. In this region the
top production comprises the most important background.
Usual ways to reduce the background are lepton isolation, lepton pair mass
constraints around $Z$ mass and
a lepton detection threshold  around
$P_T^{\ell}
\simeq 10$ GeV \cite{lhc}. In this paper we do not impose any cut on physical
variables, but we simply choose a set of physical observables
whose values represent the entries of a neural classifier,
since we expect that, by an appropriate choice of such observables,
the discrimination should automatically occur as a result of the neural
classifier.

Before discussing the variables let us examine
the background in  more detail. Besides $t \bar t$ production
followed by top semileptonic decay to bottom and $b$ semileptonic decay, one
expects other sources of background, most notably
$Zb \bar b$; however, as shown by the LHC study groups, this
process is expected to contribute by around only one third of $t \bar t \to
4 \mu$ to the cross section;  therefore we shall neglect it at this stage
because in this letter we are more interested in
a study of the relative performance of the NN and the multivariate analysis
rather than a comprehensive study of the background. In any event we do not
expect
a significant change of the results, should these other minor
backgrounds be included.

On the other hand  several other background processes
become important if one does not impose cuts on lepton transverse
momenta; to take into account them, together with the dominant
$t \bar t \to
4 \mu X$,
we shall include as background processes all the events where four
muons are in the final state and a $t \bar t$ pair has been produced,
without forcing their decay. We have checked, by simulating the events
by the Pythia Montecarlo\cite{pyt}, that these background events produce
$\sigma \times BR \simeq 11 pb$, which is larger by a factor of about $25$ as
compared to $\sigma \times BR$ for $pp \to t \bar t X \to 4 \mu X$,
choosing a top mass of 130 GeV \cite{ffr}.

Let us now list the physical observables we have used for
discrimination between background and signal.  We have considered
10 observables, that are:
\begin{description}
\item[$X_1$) - $X_4$)] the transverse momentum of the four muons. The two
 $\mu^+$ and the two $\mu^-$  can be ordered according to
their energies.
As expected, the distributions of these variables for background
events, as simulated by the Pythia Montecarlo, show a maximum close
to zero, while the
signal distributions show a peak around 25 -- 50 GeV,
\par\noindent
\item[$X_5$) - $X_8$)]The  invariant masses of the four
different $\mu^+ \mu^-$ pairs. Also these pairs can be ordered according
to the lepton energies.
These distributions for signal events show a peak around the $Z^0$ mass
which are absent for the background. The peaks arise from events where two
muons come
from the real $Z^0$; they are present in all the 4 variables since the
ordering based on the
energy mixes in part the muons coming from the two $Z^0$,
\par\noindent
\item[ $X_9$)] The four muons invariant mass,
\item[$X_{10}$) ] The hadron multiplicity.
\end{description}

This choice of variables is mainly based on kinematical considerations and
should be considered as {\it minimal}, since we expect that
other dynamical
variables, besides the hadron multiplicity, can improve the performance
of the network. We shall come back to this point later on.

The physical observables $X_j$ discussed above, once normalized to the interval
$[0 - 1]$, become the inputs $x_j \; (j=1,...n)$
of our neural network classifier. We
employ the most common architecture used for high energy applications (see,
e.g. \cite{nn}), i.e.
the {\it feed-forward} neural network; in our case it
comprises one input layer with
 $n=10$ neurons $x_j$, one layer with $2n+1$ hidden neurons $z_j$
and one output unit $y$. We employ the {\it backpropagation} algorithm
\cite{rum} to train the network. The events are divided into two sets,
the training and the testing set. Each event $p$ of the training set
consists of the array $x_j$ of the input variables and the value $y$ of the
output neuron ($y=0$ or $1$ for the event describing the signal, i.e.
Higgs production, or the background, respectively). At each time
step and for any pattern $p$ in the training set, the algorithm
modifies the synaptic
couplings giving the strength of the interaction between the hidden
layer and the output neuron:
\be
W_{i} \; \rightarrow \; W_{i} \; + \; \Delta W^{(p)}_{i}
\ee
\par\noindent
with
\be
\Delta W^{(p)}_{i}  \; =\; -\lambda {\partial E^{(p)} \over
\partial W_{i} } \; + \; \alpha \Delta W^{(old)}_{i} \; ,
\ee
\par\noindent
where
\be
E^{(p)}= {1 \over 2} (y^{(p)} -t^{(p)})^2
\ee

In the previous equation,
for any pattern $p$ in the training set, $t^{(p)}$ is
the expected {\it target} ($t^{(p)}= 0$ or $1$, for signal and
background event respectively) and $y$ is given by:
\be
y \; = \; g(u, \theta) \;
\ee
\par\noindent
where the transfer function $g(u,\theta)$ is as follows:
\be
g(u,\theta) \; = \; {1 \over 1 + \exp (- {u - \theta \over T })}
\ee
and $u$ is given in terms of the hidden variables by $u \; =
\sum_{l=1}^{2n+1} W_{l} \, z_l$.
Similar relations hold among the hidden variables $z_k$ and the input
neurons $x_j$, so that
the repeated application of Eqs. (2.1) and (2.2)
fixes the couplings $W_{lk}$  between the hidden neurons $z_k$ and the
input neurons $x_l$ as well. In our simulations we use
the values $\alpha =0.9$, $\lambda=1$ and $T=1$ for the network parameters.

\resection{Results}
Our simulations have been obtained by
the Pythia Montecarlo Code \cite{pyt}. We have considered two masses
of the Higgs particle: one below $2 M_Z$ i.e. $150$ GeV, and
one just above, i.e. $200$ GeV. The
top quark mass has been put equal to
130 GeV (increasing $m_t$ improves the
results since it reduces the background).
The simulated events have been divided into two sets,
the training set, used by the network to learn, and the testing set,
used to test the performance of the NN.
The training set consists of an equal number $N \,=\,  5,000 $
of background and signal events:
we have checked that the results are stable against changes of $N$.
On the other hand in the testing set the populations of the two samples,
background and signal,
have been taken different; as a matter of fact, we have considered
2,000 signal $pp \to H X \to 4 \mu X$
events, independently of the Higgs mass, while
one has to take $1.1 \times 10^7$ and $4.3 \times 10^6$ background events
for the two cases of $M_H \, = \, 150$ and $200$ GeV respectively.
The ratios between signal and background cross section
that we use are computed in the Standard Model by the Pythia Montecarlo.

If the sample is statistically significant one may consider a smaller set of
background events and rescale the final results (i.e. the number of
misinterpreted background events) according to the predictions of the
Standard Model. We have checked that this procedure works already with 50,000
background events.

We have considered four different cases in our simulations.
In the first case the 10 variables have been
used with no cuts; the number of simulated events quoted above refers
to this case. In the second case, in
order to increase the ratio signal/background, we have considered
only muons with the $4 \, \mu$ invariant mass
in the range $M_H \,\pm 10$ GeV. In these simulations we have obtained
slightly
worse results, i.e. the discrimination seems more
difficult; moreover, in this case, the training phase
lasts longer.

In order to determine the most suitable variables for this kind of study,
we have also repeated the analysis by using only five variables. We have
considered two of these cases, one with $ x_1,\, x_2,\, x_3,\, x_4$ and
$x_9$ as input variables and the other with input variables
$x_5,\, x_6, \, x_7,\, x_8,\, $ and $x_9$.
In both these cases the results are slightly worse as compared
to the choice of 10 variables. Therefore, in the evaluation of
the performance of the NN we shall refer
to the first of the four cases we have discussed, i.e. 10 variables
with no cut on $M_{\mu \mu \mu \mu}$.

The performance of the NN classifier can
be assessed by introducing two
variables: the purity ($P$) and the efficiency ($\eta$) defined as follows:
\be
P \, =\, \frac{N^a_H}{N^a_H + N^a_B}
\ee
and
\be
\eta \, = \, \frac{N^a_H}{N_H}
\ee
where $N_H$ is the total number of Higgs events in the testing sample, $N^a_H$
is the total number of the accepted
(i.e.
correctly identified)
Higgs events and
$N^a_B$ is the total number of the
accepted background events,
i.e. events that are incorrectly identified as
Higgs events.

One can increase the purity decreasing the efficiency
by introducing a
threshold parameter $l \epsilon [0,1]$ as follows.
The range of values of the output
neuron $y^{(p)}$ in the testing phase is divided into
the subintervals: $I_1 =
[0,l]$ and $I_2 = ]l,1]$, so that if $y^{(p)} \epsilon
I_1$ (respectively  $y^{(p)} \epsilon I_2$) the event is
classified as signal  (respectively: background).
Clearly, by taking $l$ sufficiently small, one increases purity. For example
at $M_H \, = \, 150$ GeV one obtains $P=0.1$ for $l \approx 10^{-3}$ and
$P=0.25$ for $l \approx 0.5 \times 10^{-4}$.

Our results are reported in Fig. 1, which shows that,
as expected, the case with $M_H \, = \, 200$ GeV is certainly
more favourable than the case with $M_H \, = \, 150$ GeV.  Fig. 1 also shows
that in both cases one can reach appreciable values of purity,
even though, especially for low Higgs mass, the reduction of efficiency
is relevant.

Let us now compare these results with the maximum
likelihood method.
First of all, as a general comment, we observe that the neural network
is more flexible than the multivariate analysis, since in the former case
by an appropriate choice of the parameter $l$ one can increase the purity
without limitations, at least in principle. This flexibility is absent
in the multivariate analysis because this method only uses averages
and not a uniform fit to the data as the NN does.
When the comparison is possible, i.e. for values of
the efficiency larger then $30\%$, the maximum likelihood method
gives results that are significantly worse. By way of example, at
$M_H \, = \, 150$ GeV with an efficiency of $85\%$
the traditional method gives a purity of 0.02 (a factor of 3 worse
than the NN result of Fig. 1);
with an efficiency of $99\%$
the traditional method gives a purity of 0.01, again worse than NN.
Similar results are obtained with $M_H \, = 200$ GeV: for example,
the
efficiency of $34\%$ corresponds, with
the traditional method, to a  purity of 0.07 , while NN
gives a purity of roughly 0.35 at the same value of efficiency.

\resection{Conclusions}

Our results show that NN can be of some help in the difficult task of
discriminating background events from the signal in the Higgs search
at the future Large Hadron Collider to be built at CERN. We have
proved this by considering one particular Higgs decay channel ($H \to 4 \mu$)
in the Higgs mass range (150-200) GeV and including the
most relevant background. We are conscious of the limits of the present
analysis: for example other sources of background should be included and
different Montecarlo's might be employed to test the independence
of the results from the theoretical inputs. Moreover
other global variables, similar to
the hadron multiplicity and sensitive to
the infrared structure of the QCD radiation could be introduced,
even though the use of preprocessed
observables might limit the whole range
of possibilities of the neural trigger. We plan to perform these
analysis in a subsequent paper. On the other hand, from our experience on
a similar subject \cite{mar} we do not expect a dependence of the results
on the architecture of the neural network. We feel,
therefore, that we have correctly addressed the main point
i.e. the comparison between the neural network
and the usual multivariate analysis
based on the maximum likelihood method.
Our results show that NN
compare favourably with the traditional
statistical analysis.
Needless to say that NN have another
clear advantage over traditional statistical methods, since they
can support a high degree of parallelism and could be used for
on-line analysis of the experimental data. Therefore their
use in the future LHC experiments should be seriously considered
and thoroughly investigated.
\vskip 1cm

{\bf Acknowledgements.} We wish to thank G. Marchesini for several
helpful comments on the subject of this work and M.C. Cousinou, S. Basa and
C.Marangi for useful discussions.

\newpage

\begin{center}
  \begin{Large}
  \begin{bf}
  Figure Caption
  \end{bf}
  \end{Large}
\end{center}
  \vspace{5mm}
\begin{description}
\item [Fig. 1] The purity $P$ versus the Higgs efficiency
$\eta$ for two different sets of data: $M_H \, = \, 150$ and
200 GeV (lower and upper line respectively).
\end{description}
\end{document}